\documentclass[11pt]{article}
\usepackage[top=2cm,left=2cm,right=2cm]{geometry}
\usepackage[numbers,square]{natbib}

\usepackage[italian,english]{babel}
\usepackage{epsfig}
\usepackage{hyperref}
\usepackage{amssymb}
\usepackage{amsfonts}
\usepackage{epsf}
\usepackage{rotating}
\usepackage{graphicx}
\usepackage{amsmath}
\usepackage{fancyhdr}
\usepackage{lineno}
\usepackage{subfigure}
\usepackage{babel}
\usepackage{graphics}
\usepackage{geometry}
\usepackage{pstricks}
\usepackage{color}

\makeatletter
\def\cleardoublepage{\clearpage\if@twoside
\ifodd\c@page
\else\hbox{}\thispagestyle{empty}\newpage
\if@twocolumn\hbox{}\newpage\fi\fi\fi}
\makeatother

\newcommand{\bea}{\begin{eqnarray}}
\newcommand{\eea}{\end{eqnarray}}

\def\nn{\nonumber}

\def\beq{\begin{equation}}
\def\eeq{\end{equation}}

\def\ba{\begin{eqnarray}}
\def\ea{\end{eqnarray}}

\newcommand{\beqa}{\begin{eqnarray}}
\newcommand{\eeqa}{\end{eqnarray}}

\newcommand{\veps}{\varepsilon}
\newcommand{\la}{\lambda}

\begin{document}
\begin{center}
\vspace{4.cm}

{\bf \large Comments on Anomaly Cancellations 
by Pole Subtractions\\ 
and Ghost Instabilities with Gravity}
\vspace{1cm}

{\bf Roberta Armillis $\footnote{roberta.armillis@le.infn.it}^{(a,b)}$, Claudio Corian\`{o} $\footnote{claudio.coriano@unisalento.it}^{(a)}$, Luigi Delle Rose$\footnote{luigi.dellerose@le.infn.it}^{(a)}$ and A. R. Fazio$\footnote{arfazio@unal.edu.co}^{(c)}$}

\vspace{1cm}

{\it $^{(a)}$Dipartimento di Fisica, Universit\`{a} del Salento \\
and  INFN-Lecce, Via Arnesano 73100, Lecce, Italy
}\\
\vspace{0.5cm}
{\it $^{(b)}$Physics Division, School of Technology, \\Aristotle University of Thessaloniki,
 Thessaloniki 54124, Greece\\}
\vspace{0.5cm}
{\it $^{(c)}$Departamento de Fisica, Universidad Nacional de Colombia\\
 Ciudad Universitaria, Bogot$\acute{a}$, D.C. Colombia\\}

\vspace{.5cm}
\begin{abstract}
We investigate some aspects of anomaly cancellation realized by the subtraction of an anomaly pole, stressing on some of its properties in superspace.  In a local formulation these subtractions can be described in terms of a physical scalar, an axion and related ghosts. They appear to be necessary for the unitarization of the theory in the ultraviolet, but they may generate an infrared instability of the corresponding effective action, signalled by ghost condensation. In particular the subtraction of the superanomaly multiplet by a pole in superspace is of dubious significance, due to the different nature of the chiral and conformal anomalies. In turn, this may set more stringent constraints on the coupling of supersymmetric theories to gravity.

\end{abstract}
\end{center}
\newpage
\section{Introduction} 

Two different approaches appear in the description of the mechanism of anomaly cancellation, involving either a counterterm in the form of a pole subtraction \cite{LopesCardoso:1991zt,LopesCardoso:1992yd},  or a Wess-Zumino term  (see for instance \cite{Elvang:2006jk}). This goes under the name - rather generically - of the 
Green-Schwarz mechanism (GS) in four dimensional field theory. 
These two forms of the mechanism at the level of the 
1-particle irreducible (1PI) effective action are, obviously, not equivalent, and the issue of their completeness, from a field theory point of view, is still open. For instance, axionic shift symmetries, which are present 
in some formulations of gauged supergravities, have been investigated using a Wess-Zumino approach \cite{DeRydt:2007vg,Zagermann:2008gy}. On the other hand, the subtraction of the anomaly pole in superspace - which is the one that we will mostly address in this note - has also been introduced as a possible way to give consistency to the effective action, in the presence of quantum anomalies. At the same time, a large amount of work along the years has addressed the problem of anomaly cancellation in matter-coupled supergravities using, at least in some cases, the subtraction mechanism. These studies have been and are focused on 
the role of K\"ahler and sigma model anomalies \cite{LopesCardoso:1992yd,Derendinger:1991hq,Butter:2009je,Freedman:2005up} 
and on their implications in anomaly-mediated supersymmetry breaking \cite{Bagger:1999rd}.

\subsection{Open issues}
The goal of this note is to stress on some 
(and unique) features of this subtraction from a perturbative perspective, in particular on the issues left open - at field theory level - and which have not yet found a satisfactory answer.  In particular, we point out that there are, indeed,  two challenges to the understanding of this mechanism in field theory. They are related  1) to the presence of ghosts in the spectrum of anomalous theories after the subtraction and 2) to the question whether a simple pole subtraction can actually erase the trace anomaly, in case also this needs to be cancelled. This second point is rather subtle since in supergravities the gauging combines several different symmetries, by requiring the invariance of the complete action under a combination of scaling symmetries (super-Weyl) together with ordinary K\"ahler transformations in addition to a $U(1)_R$ gauge symmetry.

A third issue concerns the relation between anomaly induced actions, which are derived by a solution of the anomaly 
equation, and the complete perturbative action obtained from a direct (and complete) diagrammatic approach. Both methods determine effective actions which are characterized by anomaly poles, the second approach being, obviously, more complete.  
Explicit computations, in fact, allow to understand the significance of the anomaly poles also as specific ultraviolet (UV) contributions, emerging from the perturbative expansion in the large energy limit. This point, as we are going to explain below, allows to put into the right context the meaning of the subtraction mechanism, which should be part of a UV completion. 

 All these issues have some implications for supersymmetric Yang Mills theories when these are coupled to 
conformal supergravity or to the various (old and new) multiplets of Poincare supergravities, due to the emergence of an infrared instability at perturbative level, induced by the mechanism. This can be identified by a direct analysis of the Coleman-Weinberg potential of the corrected theory, which shows the presence at 1-loop level of a ghost condensate. 

Therefore, a true understanding of the mechanism of 
anomaly mediation and/or cancellation, to be significant at phenomenological level, has to address the role of the axion-ghost system and of the scalar-ghost system which, as we are going to explain, are introduced by these subtractions.
Our simplified analysis has the role to stress the essential features of the pole subtraction, using very simple examples, but coming 
to conclusions which are, in fact, quite general. 
As we are going to show, much of the problem arises due to the nature of these pole counterterms in perturbation theory. The lifting of this approach to superspace, while necessary, complicates considerably the matter, especially since chiral gauge anomalies and trace anomalies may be jointly involved in the cancellation. This may happen if the K\"ahler symmetry has physical significance and needs to be preserved \cite{Freedman:2005up}. 

\section{Removing the chiral gauge anomaly by an axion or by a pole subtraction}
The simplest Lagrangians that in field theory realize the Wess-Zumino version of the mechanism can be written down quite straightforwardly, starting, for instance, with a single anomalous 
$U(1)_B$   model. It is defined as 
\begin{eqnarray}
&&\emph{L}=\bar{\psi}(i{\partial}+g{B}\gamma_5)\psi-\frac{1}{4}F^2_B + \langle \Delta_{BBB} BBB\rangle + c_1 \frac{b}{M} F_B\wedge F_B
\label{firstAN}
\eeqa
and contains one chiral fermion, which indeed introduces an anomaly at quantum level. A discussion of this action is given in 
\cite{Preskill:1990fr}. We have included in its structure the 
$ \langle \Delta_{BBB} BBB\rangle$ interaction, which represents the contribution from the triangle diagram \cite{Coriano:2007fw}. 
We can fix the counterterm $c_1$ from the requirement of gauge invariance, balancing the anomalous variation of the anomaly diagram with the variation of the axion counterterm. The axion undergoes a local shift under a gauge transformation
\beq
\delta b = M \theta_B(x) \qquad \qquad \delta B_\mu= \partial_\mu \theta_B(x)
\label{shifts}
  \eeq
where $\theta_B(x)$ parameterizes a gauge transformation. The Lagrangian implements in a simple form the GS mechanism 
(via an asymptotic axion $b$) and is obviously generalizable to supersymmetry via a shifting supermultiplet (see for instance \cite{Elvang:2006jk} and \cite{Coriano:2010ws} for a theoretical and phenomenological discussions in the supersymmetric case). As we have already mentioned, there is no equivalence between the pole subtraction mechanism and  the Wess-Zumino counterterm, and these approaches are sometime not clearly distinguished in the literature. This difference, at the level of the 1-particle irreducible effective action, is indeed substantial. 

The model Lagrangian introduced in (\ref{firstAN}) has some pitfalls, the first of them being the absence of a kinetic term for the axion. We can try to avoid the problem by introducing a kinetic term in a gauge invariant form. There is only one possibility, the St\"uckelberg mass term, obtaining the modified action

\begin{eqnarray}
&&\emph{L}=\bar{\psi}(i{\partial}+g{B}\gamma_5)\psi-\frac{1}{4}F^2_B + \langle \Delta_{BBB} BBB\rangle + c_1 \frac{b}{M} F_B\wedge F_B + \frac{1}{2} \left(\partial_\mu b - M B_\mu\right)^2.
\label{WZ1}
\eeqa

This Lagrangian has a typical $M b\, \partial B$ interaction that one could try to remove via a gauge fixing. In fact, one can do so and investigate the behaviour of the perturbative expansion in such a gauge (of $R_\xi$ type). These studies have been performed in 
\cite{Coriano:2007fw}.  The theory describes consistently the mechanism of anomaly cancellation up to a certain scale, which is essentially the St\"uckelberg Mass $M$, since there is, indeed, a 
unitarity bound. There is a second limitation of this type of action, coming directly from gauge invariance. In fact one could 
choose a gauge in which $b$ is set to vanish, and the theory would turn out to be equivalent to a massive Yang Mills theory coupled to a chiral fermion. For this reason, this action should necessarily be viewed as an approximate description of a more general one. This could be deduced starting from an anomaly free theory and decoupling even a single chiral fermion from the functional integral \cite{Coriano:2009zh}. It has been shown that the effective action obtained by this decoupling is indeed corrected by an infinite number of higher dimensional operators. In this respect, the Lagrangian given in (\ref{WZ1}) has a unitary completion, at least in a field theory sense. Notice that $b$ can be thought of as the phase of an extra Higgs field (complex scalar) having decoupled its modulus. 
 For this reason, Lagrangians of this type are sufficient to describe the leading behaviour of the effective action in a $1/M$ expansion.  

A second version of the mechanism is described instead by the second (nonlocal) Lagrangian
\begin{eqnarray}
&&\emph{L}=\bar{\psi}(i{\partial}+g{B}\gamma_5)\psi-\frac{1}{4}F^2_B + \langle \Delta_{BBB} BBB\rangle + c_2 \partial B \frac{1}{\square} 
F_B\tilde{F}_B
\label{WZ2}
\eeqa
where the term $\partial B \frac{1}{\square} F_B\tilde{F}_B$ is the anomaly pole. 
 It does not take much to realize that the cancellations corresponding to (\ref{WZ1}) and (\ref{WZ2}) allow to restore gauge invariance of the effective action. In general, extra counterterms can also be added to these types of actions in the presence of at least two gauge simmetries, in the form of Chern-Simons (CS) interactions. In the case that we consider the only possible 
 anomaly is the consistent one, given the symmetry. For all practical purposes, CS interactions simply allow to re-distribute the partial anomalies ($a_i$) on a given leg of a diagram, keeping their sum fixed ($a_1+a_2+a_3=a_n$). In the case of a theory with two U(1)'s 
 (e.g. $U(1)_A \times U(1)_B$) with $A$ vector-like and $B$ axial-vector-like, terms such as ($AB\wedge F_B, A B\wedge F_A$) allow to move from the consistent to the covariant form of the anomaly. In any case, the discussion of CS interactions is not relevant for our goals and it will be omitted. 

This second version of the mechanism, realized via  (\ref{WZ2}), introduces one additional degree of freedom compared to (\ref{WZ1}). As we are going to show, this extra degree of freedom is an anomaly ghost. In fact, the Lagrangian (\ref{WZ2}) admits a different (local) formulation, now in terms of two extra pseudoscalars  
of the form 
\beqa
\mathcal{L} &=& \overline{\psi} \left( i \not{\partial}  + e \not{B} \gamma_5\right)\psi - \frac{1}{4} F_B^2  +\langle \Delta_{BBB}BBB\rangle + c_3 F_B\wedge F_B ( a + b) \nonumber \\
&& + \frac{1}{2}  \left( \partial_\mu b - M_1 B_\mu\right)^2 -
\frac{1}{2} \left( \partial_\mu a - M_1 B_\mu\right)^2,
\label{fedeq}
\eeqa
where both $a$ and $b$ shift as in (\ref{shifts}). The equivalence between (\ref{WZ2}) and (\ref{fedeq}) can be proven directly from the functional integral, integrating out both $a$ and $b$, which gives two gaussian integrations. Notice that $b$ has a positive kinetic term and $a$ is ghost-like. 

There is a third equivalent formulation of the same action (\ref{fedeq}) which can be defined with the inclusion of a kinetic mixing between the two pseudoscalars. This has been given for QED (with a single fermion) coupled to an external axial-vector field $\mathcal{B}_\mu$ \cite{Giannotti:2008cv} and takes the form 

\beq
\mathcal{L}=\partial_\mu\eta \partial^\mu\chi - \chi \partial \mathcal{B} + \frac{e^2}{8 \pi^2}\eta F \tilde{F},
\label{GM}
\eeq
where $F$ is the field strength of the photon $A_\mu$ while $\mathcal{B}_\mu$ takes the role of a source. An anomaly pole is indeed induced by the $\mathcal{B} AA$ anomaly vertex.  It is quite straightforward to relate (\ref{fedeq}) and (\ref{GM}). This can be obtained by the field redefinitions
\beqa
\eta &=& \frac{(a + b)}{M}, \nonumber \\
\chi &=& M(a - b),
\label{changes}
\eeqa
showing that indeed a mixing term is equivalent to the presence of either an anomaly pole or to two pseudoscalars in the spectrum of the theory, one of them being a ghost. It is obvious that the pole subtraction in superspace does exactly the same thing, in a rather unobvious way. 

\subsection{The anomaly pole and the trace anomaly}
The appearance of an anomaly pole in the perturbative expansion is not limited to the chiral anomaly. To clarify this point, let's denote with $k$ the incoming momentum of the anomalous gauge current or of the graviton and with $p$ and $q$ the outgoing momenta of the two vector gauge bosons.

Similar singularities appear in explicit computations of the correlation functions for the trace anomaly in the absence of any second scale in the loop, involving one insertion of the energy momentum tensor ($T$) on 2-point functions of gauge fields $(VV')$,  the $TVV'$ correlator. By a second scale we refer either to a fermion mass term $m$ in the anomaly loop, or to any of the two virtualities $s_1$ and $s_2$ ($s_1\equiv p^2$, $s_2\equiv q^2$) of the two gauge currents. With the term "first scale" in the loop we refer to the virtuality of the graviton $s$ ($s\equiv k^2$), or, in the case of the chiral anomaly, the virtuality of the axial-vector current.
This is the scale that as $s$ goes to zero (with $k^\mu\to 0$, soft infrared (IR) limit) or as $s$ goes to infinity (i.e. $k^\mu$ goes to infinity with a large invariant mass) controls the effects of the anomaly on the trilinear vertex.
In fact the 
$TVV'$ correlator takes a role quite similar to that of the corresponding $AVV$ diagram of the chiral gauge anomaly. Surprisingly, this correlator 
has never been computed  explicitly until recently in QED, QCD and the Standard Model. In the case of QED, for instance, the effective action takes the form  
\cite{Giannotti:2008cv} \cite{Armillis:2009pq, Armillis:2010qk}
  \beq
S_{anom}[g,A]  \rightarrow  -\frac{c}{6}\int d^4x\sqrt{-g}\int d^4x'\sqrt{-g'}\, R_x
\, \square^{-1}_{x,x'}\, [F_{\alpha\beta}F^{\alpha\beta}]_{x'}\,,
\label{SSimple}
 \eeq
($c= - e^2/(24 \pi^2)$) which is valid to first order in the fluctuation of the metric around a flat background, denoted as $h_{\mu\nu}$
\beq
g_{\mu\nu}= \eta_{\mu\nu} +\kappa h_{\mu\nu}, \qquad\qquad \kappa=\sqrt{16 \pi G_N},
\eeq
with $G_N$ being the 4-dimensional Newton's constant. The pole emerges from a single form factor evaluated in momentum space. 
If we denote with $\Gamma_{\mu\nu\alpha\beta}\equiv \langle T_{\mu\nu} V_{\alpha} V_{\beta}\rangle$ the correlation function responsible for the trace anomaly, this takes the form  
\beq
\Gamma^{\mu\nu\alpha\beta}\sim \frac{1}{k^2}( g^{\mu\nu}k^2 - k^\mu k^{\nu})u^{\alpha\beta}(p,q) + ...
\eeq
where $u^{\alpha\beta}(p,q)$ is a tensor structure obtained by functional differentiation of the $FF$ term of the trace anomaly Fourier transformed to momentum space,

\beq
u^{\alpha\beta}(p,q)=-\frac{1}{4} \int d^4 x d^4 y e^{i p\cdot x + i q\cdot y}\frac{\delta^2 F_{\mu\nu}F^{\mu\nu}}{\delta V_\alpha(x) \delta V_{\beta}(y)}.
\eeq

 The ellypsis refer to terms which are traceless. This relation is the analogous of the anomaly pole expression 
\beq
\Delta_{AVV}^{\lambda\mu\nu}=a_n \frac{k^\lambda}{k^2} \epsilon^{\mu\nu\alpha\beta}p_\alpha q_\beta + ...
\eeq
for the chiral anomaly, with $a_n$ being the anomaly. The pole structure above is usually called a Dolgov-Zakharov pole (DZ), which is IR coupled only in the absence of any second scale in an anomaly diagram. It is important to remark that 
only in this case (i.e. for two on shell vector lines and massless particles in the loop) the cancellation between an anomaly diagram and the subtraction countertem  is identical. There is no identical cancellation under any other circumstance. For this obvious reason, in the presence of any second scale in the anomaly loop, 
the anomaly cancellation mechanism amounts to an "oversubtraction". 

The meaning of this last term can be clarified quite simply.
In fact we just recall that in the case of the chiral anomaly, the pole subtraction can be absorbed into a redefiniton of the anomaly vertex - this is not the case for the Wess-Zumino cancellation with a single axion ($b$)  \cite{Armillis:2008bg} - which now satisfies regular Ward identities (i.e. non anomalous) on each of its three external legs.  This redefined vertex, however, now has a pole which is infrared coupled for {\em any} virtuality of the external vector lines, a feature which 
is unique among all the known vertices in local quantum field theory and, in particular, in the Standard Model. We will come back to this point in the next sections, trying to address the issue in the case of the chiral anomaly vertex.

As in the case of the chiral anomaly pole, also for the trace anomaly two auxiliary fields allow to re-express in a local form 
the corresponding nonlocal action (\ref{SSimple}) which takes the form \cite{Giannotti:2008cv}
\beq
S_{anom} [g,A;\varphi,\psi'] =  \int\,d^4x\,\sqrt{-g}
\left[ -\psi'\square\,\varphi - \frac{R}{3}\, \psi'  + \frac{c}{2} F_{\alpha\beta}F^{\alpha\beta} \varphi\right]\,,
\label{effact}
\eeq
where $\phi$ and $\psi^\prime$ are auxiliary scalar fields. Also in this case one can perform the same changes of variables as in (\ref{changes}) and remove the kinetic mixing form this Lagrangian. Notice that the two auxiliary fields, in this case, are scalars.  One of the two degrees of freedom is indeed a ghost. It is then clear that the subtraction of a anomaly pole induces into the theory some ghosts which are supposed to cancel those present in the trilinear anomalous vertices. As we are going to show, simple arguments in perturbation theory show that as soon as \ref{WZ2} is used in the computation of quantum corrections, one discovers the presence of an infrared instability. For this we need to use the local version of (\ref{WZ2}), but before moving to that discussion we briefly 
comment on some of the main features of a pole subtraction in superspace.

\subsection{The superconformal case and the gauging to gravity}
Several puzzles emerge as soon as we put together the pieces of our previous discussion and frame it into a supersymmetric context (see \cite{Chaichian:2000wr, Dienes:2009td} for an overview).  

When we come to analyze a super Yang-Mills theory, the trace anomaly, the gamma-trace of the supersymmetric current and the anomaly of the $U(1)_R$ current are part of the same anomaly supermultiplet $(T^\mu_\mu, \gamma \cdot s, \partial J_5)$ 
\cite{Ferrara:1974pz}. In this case the supermultiplet describes the radiative breaking of the superconformal symmetry. In particular, the presence of an anomaly pole for the axial-vector $U(1)_R$ global current indeed implies that a similar pole should appear in the  correlation functions involving the insertion of either an energy-momentum tensor - or of the supersymmetric current - on two vector currents. This result is necessary for a consistent formulation of the anomaly-free effective action in superspace. Indeed, explicit computations support this picture to lowest order in the case of the trace anomaly, being obviously true (and to all orders) for the $U(1)_R$ anomaly.

The gauging of such an anomaly multiplet to gravity,
for instance via a conformal multiplet $(g_{\mu\nu}, \psi_\mu, B_\mu)$ containing a graviton, a gravitino and an axial-vector gauge field, indeed produces an anomaly. In this case the energy momentum tensor couples to gravity $(g_{\mu\nu})$, the supersymmetric current couples to the gravitino background 
$(\psi_\mu)$ and the anomalous $U(1)_R$ current couples to the axial-vector gauge boson $B_\mu$. Diffeomorphism invariance 
gives the standard conservation conditions for $T^\mu_\nu$ and the spinor current $s_\mu$ ($\nabla_\mu T^{\mu\nu}=0, \nabla_\mu s^\mu=0$), but the super-Weyl and $U(1)_R$ symmetry of the theory ($(T^\mu_\mu=0, \gamma \cdot s=0, \partial J_5=0)$ are radiatively broken (see also \cite{Freedman:1976uk, Castano:1995ci,Chamseddine:1995gb} for related studies). It is obvious that the cancellation of the superconformal anomaly can't be obtained by using a single pole in superspace, given the different nature of the chiral and trace anomalies.

Anomaly induced actions \cite{{Bagger:1999rd}} for $N=1$ matter coupled supergravities carry both the signature of the breaking of scale invariance and of 
gauge invariance under Super-Weyl-K\"ahler transformations of the effective action, as shown by the presence both of the  
1) $R\,\square^{-1} FF$ and of the 2) $\partial B\,\square^{-1} F\tilde F$ terms in the effective action, with $R$ being the scalar curvature
 \cite{{Bagger:1999rd}} \cite{LopesCardoso:1991zt}.

While the appearence of the second term is, in a way, obvious, since it is generated by the Dolgov-Zakharov (DZ) anomaly pole present in the 
$AVV$ diagram in superspace\cite{Dolgov:1971ri}, the first one is far from being obvious since its identification requires a rather involved computation of the full correlator, not carried out until recently \cite{Giannotti:2008cv, Armillis:2009pq, Armillis:2010qk}. Similar poles emerge in the same vertices of the Standard Model, so far computed in the case of the neutral currents \cite{Armillis:2010pa, Coriano:2011ti}. It is then amusing that the lifting to superspace of the DZ pole of the $U(1)_R$ current, induces a similar pole in the correlator responsible for the trace anomaly. 

It is however clear that the $R\,\square^{-1} FF$ result is just valid to lowest order ($O(G_N g^2)$) in Newton's constant $G_N$ and gauge coupling $g$. Indeed, in general, the structure of the anomaly-induced effective action for the trace anomaly is expected to be far more involved compared to the simple pole result. For instance, this action should describe the structure of the singularities of anomalous correlators with any number of insertions of the energy momentum tensor and two photons ($T^n VV$). 

For obvious reasons, explicit checks of the corresponding effective action using perturbation theory - as the number of external graviton lines grows -  becomes increasingly difficult to handle. The $TVV$ correlator is the first (leading) contribution  to this infinite  sum of correlators in which the anomalous gravitational effective action is expanded. One proposal for the effective action is due to Reigert 
\cite{Riegert:1984kt}, which has been successfully tested, so far only for the $TVV$ case, by two independent groups \cite{Giannotti:2008cv, Armillis:2009pq, Armillis:2010qk}. 

Given the presence of a quartic operator in Riegert's nonlocal action, the proof that this action contains a single pole to lowest order (in the TVV vertex), once expanded around flat space, has been given in \cite{Giannotti:2008cv} and provides the basis for the discussion of the anomalous effective action (\ref{effact}) in terms of massless auxiliary fields.

This shows that the ghost appearing in the trace anomaly is a genuine result which is extracted in two ways: 1) by integration of the anomaly and 2) by a direct perturbative computation 
using dispersion theory \cite{Giannotti:2008cv} or the complete evaluation of the diagrammatic expansion 
\cite{Armillis:2009pq, Armillis:2010qk}.

 \section{The Coleman-Weinberg potential for the corrected action}
We have clarified that a Lagrangian containing a pole counterterm shows some nontrivial features. Here we would like to remark on a consequence of the presence of an anomalous interaction of the form $\partial B \square^{-1} F\tilde{F}$ induced by an anomaly (DZ) pole. We are going to compute the effective potential of the ghost field $a$ defined in (\ref{fedeq}) using the standard Coleman-Weinberg approach. We will discover the presence of an instability in this potential, obtained after integration over all the remaining fields of the model. The instability is signalled by the presence of a ghost condensate at 1-loop level.  

Consider the gauge-fixed version of the Lagrangian in (\ref{fedeq}), with the omission of the $\langle\Delta_{BBB}BBB\rangle$ triangle term. Notice that the cancellation between the triangle contribution (the $\Delta_{BBB}$ term) and the $(a,b)$ part of the action takes place only when two of the three lines of the $B$ vertex are on shell and massless. The anomaly vertex, in fact, can be written as a symmetric combination of three $AVV$ graphs 
\beq
\Delta_{BBB} = \frac{1}{3}\left( \Delta_{AVV} +  \Delta_{VAV} + \Delta_{VVA} \right)
\eeq
by distributing the axial-vector current symmetrically on the three $B$ lines. Each of these three contributions develops a pole which is 
infrared coupled when the remaining "V" lines are on-shell. If we take any configuration which is indeed pole-like, then the $(a,b)$ part of the action should cancel the corresponding pole. It is clear that if we are away from any of these configurations, the missmatch between the anomaly and the counterterm is significant and this is at the core of the perturbative instability that one encounters in these types of actions. If we intend to compute the effective potential in the background of the ghost field ($a(x)$) we need only to trace the propagator of the $B$ field in the background of the ghost. To perform this computation at leading order in a loop expansion 
we need the Lagrangian 

\begin{eqnarray}
&&\emph{L}=\bar{\psi}(i\hat{\partial}+e\hat{B}\gamma_5)\psi-\frac{1}{4}F^2_B -\frac{(\partial_\mu B^\mu)^2}{2\alpha} + \frac{e^3}{48\pi^2M_1}F_B\wedge F_B (a+b)\nonumber\\
&&+\frac{1}{2}(\partial_\mu b -M_1 B_\mu)^2 -\frac{1}{2}(\partial_\mu a -M_1 B_\mu)^2
\label{fed}
\end{eqnarray}
where $\alpha$ is the gauge parameter and we have introduced the explicit expression of the counterterm $c_3$.

We shift the ghost field, separating the classical 
ghost background (still denoted as $a(x))$, from its quantum fluctuating part on which we will integrate, $A(x)$
\begin{equation}
a(x) \longrightarrow a(x) + A(x).
\end{equation}
Dropping the linear terms in the quantum fluctuation field $A(x)$ and taking just the quadratic part of all the quantum fields we get the quadratic Lagrangian

\begin{eqnarray}
&&\mathcal{L}_{\textrm{quad}}= \bar{\psi}i\hat{\partial}\psi -\frac{1}{4}F^2_B -\frac{(\partial_\mu B^\mu)^2}{2\alpha}+ \frac{e^3}{48\pi^2M_1}a F_B\wedge F_B \nonumber\\
&& \qquad \qquad \qquad +\frac{1}{2}(\partial b)^2 - M_1 \partial_\mu b B^\mu -\frac{1}{2}(\partial_\mu A)^2 +M_1 B_\mu \partial^\mu A,
\end{eqnarray} 
from which we can determine, after an integration by parts, the contribution which is quadratic in the anomalous gauge field $B$ 
\begin{equation}
\frac{1}{2}\int d^4 x B_\mu \Big[g^{\mu\nu} \Box -\left(1-\frac{1}{\alpha}\right)\partial_\mu\partial_\nu +\frac{e^3}{24\pi^2M_1}\,\partial_\alpha a\, \epsilon^{\mu\alpha\rho\nu}\,\partial_\rho\,\Big]B_\nu.
\end{equation}

The one loop effective action in the background of the ghost $a$ is obtained by integration over all the quantum fields in the form 
\begin{eqnarray}
&&e^{i\Gamma[A]}=\int[DA][D\psi][D\bar{\psi}][DB][Db]\times \nonumber\\
&& \qquad \qquad \exp  \{i\int d^4x \Big[\bar{\psi}i\hat{\partial}\psi-\frac{1}{4}F_B^2-\frac{(\partial_\mu B^\mu)^2}{2\alpha}+\frac{1}{2}(\partial b)^2-M_1\partial_\mu b B^\mu-\frac{1}{2}(\partial_\mu A)^2+M_1B_\mu\partial^\mu A\nonumber\\
&&\qquad \qquad \qquad \qquad +\frac{e^3}{48\pi^2M_1}F_B\wedge F_B a\Big]\Big\}.
\end{eqnarray} 
Notice, in particular, that the integration over the quantum fluctuations of the ghost field $A$ gives

\begin{eqnarray}
&&\int[DA]\exp\Big[ i\int d^4x\Big({\frac{1}{2}A\Box A-M_1\partial_\mu B^\mu A}\Big)\Big]\propto\nonumber\\
&&\exp\Big [-\frac{1}{2}\int d^4 x d^4 y\Big( M_1 \partial_\mu B^\mu (x) D_F(x-y) M_1 \partial_\nu B^\nu (y)\Big)\Big] 
\label{effect1}
\end{eqnarray}
being
\begin{equation}
D_F(x-y)=\int \frac{d^4 p}{(2\pi)^4}\frac{-i e^{ip(x-y)}}{p^2-i\epsilon}
\end{equation} 
the propagator for the quantum fluctuations of the ghost field. The integration over the axion $b$ brings in some cancellations, since 
\begin{eqnarray}
\int[Db]\exp\Big[i\int d^4x\Big({-\frac{1}{2}b\Box b + M_1\partial_\mu B^\mu b}\Big)\Big]\propto\exp\Big[-\frac{1}{2}\int d^4 x d^4 y M_1 \partial_\mu B^\mu (x) D^{1}_F(x-y) M_1 \partial_\nu B^\nu (y)\Big], \nonumber \\ 
\label{effect2}
\end{eqnarray}
where we have introduced the propagator of the axion field
\begin{equation}
D^1_F(x-y)=\int \frac{d^4 p}{(2\pi)^4}\frac{i e^{ip(x-y)}}{p^2+i\epsilon}.
\end{equation}
Notice that $D_F(x-y)+D^1_F(x-y)$ vanishes in the limit $\epsilon\rightarrow 0$, thus the integration in $A$ and $b$ eliminates the terms (\ref{effect1}) and (\ref{effect2}) from the action.  Therefore we are just left with the expression
\begin{equation}
e^{i\Gamma[a]}\propto \int [DB]\exp\Big[-\frac{1}{4}i\int d^4 x(F_B)^2 -\int d^4 x \frac{(\partial_\mu B^\mu)^2}{2\alpha}+i\frac{e^3}{24\pi^2M_1}\int d^4 x a F_B\wedge F_B \Big].
\label{effective} 
\end{equation}

 Defining
\begin{equation}
l\equiv \frac{e^3}{24\pi^2M_1}\,\,\,\,\,\,\, {\rm and }\,\,\,\phi_\alpha\equiv\partial_\alpha a,
\end{equation}
the effective action of the classical background ghost field is then given by
\begin{equation}
i\Gamma[a]=-\frac{1}{2}{\rm Tr}\log\left(\Box {g^\mu}_\nu -\left(1-\frac{1}{\alpha}\right)\partial_\mu\partial_\nu + l {\epsilon^{\mu\alpha\rho}}_\nu \phi_\alpha\partial_\rho\right)
\label{finaleffect}
\end{equation}
where the trace ${\rm Tr}$, as usual, must be taken in the functional sense.

To perform the calculation of (\ref{finaleffect}) we use the heat kernel method and define the functional determinant in 
(\ref{finaleffect}) using a $\zeta$ function regularization. We take $\phi_\alpha$ to be constant. We have

\begin{equation}
\log \det Q = -\displaystyle\lim_{s\to 0}\frac{d}{ds}\frac{\mu^{2s}}{\Gamma(s)}\int^{+\infty}_{0} dt\, t^{s-1} {\rm Tr}(e^{-tQ})
\end{equation}
with the functional trace performed in the plane wave basis\begin{equation}
{\rm Tr e^{-tQ}}=\int d^4 x\, \textrm{tr} <x|e^{-tQ}|x> =\int d^4 x \, \textrm{tr}\int \frac{d^4 k}{(2\pi)^4}e^{-ikx}e^{-tQ}e^{ikx},
\label{zita}
\end{equation}
and with $\textrm{tr}$ denoting the trace on the Lorentz indices. Further manipulations give
\begin{eqnarray}
{(e^{-ikx}e^{-tQ}e^{ikx})^\mu}_\tau 
&& ={g^\mu}_\tau e^{tk^2}\exp\left({-itl{{\epsilon^\tau}_\nu}^{\alpha\rho}\phi_\alpha k_\rho}\right)+(1-\frac{1}{\alpha})\frac{k^\mu k^\nu}{k^2} (1-e^{t k^2}),
\label{matrix}
\end{eqnarray}
where we have used the relation
\beqa
e^{-t k^\mu k_\nu}&=& g^\mu_\nu+\frac{k^\mu k_\nu}{k^2}(e^{-t k^2}-1) \nonumber \\
k^\mu k_\tau e^{-itl{{\epsilon^\tau}_\nu}^{\alpha\rho}\phi_\alpha k_\rho}&=& k^\mu k_\nu. 
\eeqa
We need to consider in (\ref{matrix}) just the $\phi$-dependent part. As usual, the Coleman-Weinberg potential is gauge-dependent. 
In this case the dependence on the gauge-fixing parameter $\alpha$ can be assimilated to the constant terms.
The functional trace receives contributions only from the terms with $n$ even, and after some manipulations we obtain 
\begin{eqnarray}
tr (e^{-ikx}e^{-tQ}e^{ikx})
&=& -2e^{t k^2}\cosh{\,tl\sqrt{k^2\phi^2-(k\cdot\phi)^2}}+{\rm const}.
\label{traza}
\end{eqnarray}
Inserting (\ref{traza}) into (\ref{zita}) we get, apart from a constant factor of infinite volume, the expression of the trace 
\begin{equation}
\textrm{Tr} e^{-t Q} \sim -2\int\frac{d^4 k}{(2\pi)^4}e^{t k^2}\cosh{\,tl\sqrt{k^2\phi^2-(k\cdot\phi)^2}},
\label{integraldifficile}
\end{equation}
giving an effective potential for the background $\phi_\alpha$ of the form
\begin{equation}
V[\phi]=-\displaystyle\lim_{s\to 0}\frac{d}{ds}\frac{\mu^{2s}}{\Gamma(s)}\int^{+\infty}_{0} dt\, t^{s-1}\int\frac{d^4 k}{(2\pi)^4}e^{t k^2}\cosh{\,tl\sqrt{k^2\phi^2-(k\cdot\phi)^2}}. 
\label{effective}
\end{equation}
We can obtain the leading contribution of this effective potential by expanding the integrand in $l$, i.e. in $1/M_1$.
After performing the expansion and restoring the infinite space-time volume we obtain the effective action
\begin{equation}
S=\int{d^3 x d t }\left\{ \left(-\frac{1}{2}-\frac{3 l^2}{32 \pi^2}\right)(\partial a)^2 + \frac{5 l^4}{256 \pi^2}(\partial a)^4\right\}
\end{equation}
which obviously can be rewritten as
\begin{equation}
S=\int{d^3 x d t }\left\{-\frac{1}{2}(\partial a)^2 + \frac{5 l^4}{256 \pi^2}(\partial a)^4\right\}.
\label{effectivaction}
\end{equation}
Notice that the polynomial in the integrand 
\begin{equation}
P(\phi)=-\frac{1}{2}\phi^2+ \frac{5 l^4}{256 \pi^2}\phi^4
\label{polynomium}
\end{equation}
has a minimum at
\begin{equation}
\bar{\phi}^2\sim\frac{1}{l^2}\sqrt{\frac{128\pi^2}{5}}>0.
\end{equation}
To investigate the character of the minimum and of the fluctuations around this minimum, we select a time-like frame, where the background takes the form
\begin{equation}
\bar{a}=\bar{\phi}t.
\end{equation}
Let's now consider small fluctuactions around this configuration of minimum, denoted as $\pi$
\begin{equation}
a=\bar{\phi}t+\pi
\end{equation}
and expanding (\ref{effectivaction}) we obtain the action

\begin{equation}
S=\int d^3 x d t \left\{
\dot{\pi}^2+\sqrt{\frac{5}{128\pi^2}}l^2\dot{\pi}^3+\frac{5 l^4}{512\pi^2}\dot{\pi}^4 + \frac{5 l^4}{512\pi^2}\dot{\pi}^4 |\nabla \pi|^4-\sqrt{\frac{5}{128\pi^2}} l^2 \dot{\pi}|\nabla\pi|^2+\cdots\right\}.
\label{arka}
\end{equation}
This action has the same form as in \cite{ArkaniHamed:2003uy} (see formula (4.2)). As in this previous analysis we do not get the term $|\nabla\pi|^2$ since its coefficient is proportional to $P^\prime (\bar{\phi}) = 0$.

It is clear that Lorentz symmetry is broken, at least at 1-loop level, and shows the signal of an instability of the local model 
(\ref{WZ2}) generated in the infrared region.  Notice, in fact, that in the Coleman-Weinberg approach we are closing the gauge boson loop and we are taking the long wavelength limit of the external background ghost field. 
Finally, one should also notice that the dependence of the effective potential on $M_1$, in this approach, is recovered at higher orders. For instance, additional contributions, suppressed by $1/M_1^2$, are obtained by the insertion of the self-energy of the anomalous gauge boson on the lowest order contribution (the gauge boson loop). 

Unfortunately, the loop expansion becomes increasingly complex at higher orders, due to the non-covariant form of the propagator, and increasingly unmanageable. These features of actions containing Wess-Zumino terms have been studied in the past with similar results \cite{Andrianov:1998ay} (for studies in D-brane theories see \cite{Mavromatos:2010ar}).

Obviously, this leaves wide open the question about the possible completion of these theories in the UV. In turn, the completion would allow us to avoid the problem of the "oversubtraction" in the infrared while, at the same time, would guarantee a smooth completion of the mechanism in the UV. 

In the absence of a consistent formulation of this completion derivable in field theory (from the string theory side), one could get some hints about the structure of the corrections in the field theory case. For this reason we turn back, once more, to discuss the structure of the anomaly vertex. We will be using a special representation of the diagram in which the pole is separated from the remaining contributions under any kinematical configuration. We will attribute the pole cancellation and the unitarization as being of different origins, identifying some left over terms which are necessary so to make the pole subtraction consistent at all scales.

\section{Features of an anomaly pole and oversubtractions}
Once we allow a pole solution of the anomalous Ward identities (see \cite{Armillis:2009im} for a general discussion) of a certain correlator, we need to define the kinematical range in which this solution is reproduced in perturbation theory, since explicit computations show that the tensor decompositions of anomaly diagrams are not unique. We start with the case of the $AVV$ 
diagram. For simplicity, we will still denote with $k$ the incoming momentum on the axial-vector line, and use symmetric expressions ($k_1\equiv p$, $k_2\equiv q$) for the two outgoing momenta of the vector lines. $s\equiv k^2$ denotes the virtuality of the momentum of the axial-vector current. We have the standard parameterization due to Rosenberg \cite{Rosenberg:1962pp}
\bea
\Delta_0^{\la\mu\nu} &=& A_1 (k_1, k_2) \veps [k_1,\mu,\nu,\la] + A_2 (k_1, k_2)\veps [k_2,\mu,\nu,\la] +
A_3 (k_1, k_2) \veps [k_1,k_2,\mu,\la]{k_1}^{\nu} \nonumber \\
&+&  A_4 (k_1, k_2) \veps [k_1,k_2,\mu,\la]k_2^{\nu}
+ A_5 (k_1, k_2)\veps [k_1,k_2,\nu,\la]k_1^\mu
+ A_6 (k_1, k_2) \veps [k_1,k_2,\nu,\la]k_2^\mu.\nonumber \\
\label{Ros}
\eeqa
This parameterization is not always the most convenient. For instance, if one wants to study the mechanism of pole subtraction, 
it is convenient to use Schouten's relation and re-express Rosenberg's expression in an alternative form.  
A second decomposition of the anomaly graph into longitudinal and transverse form factors \cite{Knecht:2003xy} is possible. It has been shown \cite{Armillis:2009sm} that this representation is equivalent to the Rosenberg expression \cite{Rosenberg:1962pp} (see the discussion in \cite{White:2010ay}). It takes the form   
  
  \beq
 \mathcal \, W ^{\lambda\mu\nu}= \frac{1}{8\pi^2} \left [  \mathcal \, W^{L\, \lambda\mu\nu} -  \mathcal \, W^{T\, \lambda\mu\nu} \right],
\label{long}
\eeq
where the longitudinal component
\beq
 \mathcal \, W^{L\, \lambda\mu\nu}= w_L  \, k^\lambda \veps[\mu,\nu,k_1,k_2]
\eeq
(with $w_L=- 4 i /s $) describes the anomaly pole, while the transverse contributions take the form
\beqa
\label{calw}
{  \mathcal \, W^{T}}_{\lambda\mu\nu}(k_1,k_2) &=&
w_T^{(+)}\left(k^2, k_1^2, k_2^2 \right)\,t^{(+)}_{\lambda\mu\nu}(k_1,k_2)
 +\,w_T^{(-)}\left(k^2, k_1^2,k_2^2\right)\,t^{(-)}_{\lambda\mu\nu}(k_1,k_2) \nonumber \\
 && +\,\, {\widetilde{w}}_T^{(-)}\left(k^2, k_1^2, k_2^2 \right)\,{\widetilde{t}}^{(-)}_{\lambda\mu\nu}(k_1,k_2),
 \eeqa
with the transverse tensors given by
\beqa
t^{(+)}_{\lambda\mu\nu}(k_1,k_2) &=&
k_{1\nu}\, \veps[ \mu,\la, k_1,k_2]  \,-\,
k_{2\mu}\,\veps [\nu,\la, k_1, k_2]  \,-\, (k_{1} \cdot k_2)\,\veps[\mu,\nu,\la,(k_1 - k_2)]
\nonumber\\
&& \quad\quad+ \, \frac{k_1^2 + k_2^2 - k^2}{k^2}\, \, k_\la \, \,
\veps[\mu, \nu, k_1, k_2]
\nonumber \ , \\
t^{(-)}_{\lambda\mu\nu}(k_1,k_2) &=& \left[ (k_1 - k_2)_\la \,-\, \frac{k_1^2 - k_2^2}{k^2}\,\, k_\la \right] \,\veps[\mu, \nu, k_1, k_2]
\nonumber\\
{\widetilde{t}}^{(-)}_{\lambda\mu\nu}(k_1,k_2) &=& k_{1\nu}\,\veps[ \mu,\la, k_1,k_2] \,+\,
k_{2\mu}\,\veps [\nu,\la, k_1, k_2] \,
-\, (k_{1}\cdot k_2)\,\veps[ \mu, \nu, \la, k].
\label{tensors}
\eeqa
One should notice the presence of pole-like singularities in both the $L$ and the $T$ components proportional to $s$, which clearly 
invalidate the separation as $s$ goes to zero. The presence of such singularities is also the signal that in the absence of any extra scale beside $s$, the two terms ($L/T$)  reduce to a single structure. 

To illustrate this point, let's consider in fact the case $s_1=s_2=0$. In this case the two nonzero form factors are  $w_L$ and $w_T^{(+)}$
\bea
w_L(s, 0,0) &=& w_T^{(+)}(s, 0,0) = - \frac{4i}{s}, \\
w_T^{(-)}(s, 0,0) &=& \tilde{w}_T^{(-)}(s, 0,0) = 0.
\eeqa
The only contributions to the anomaly vertex come from the longitudinal $W_L$ component and by $t^{(+)}_{\lambda\mu\nu}$, the second one being irrelevant when the two vector lines are set on-shell. Therefore, the parameterization reduces only to the longitudinal contribution, and generates, correctly, the anomaly pole. This is essentially the only case in which the pole is IR coupled, since with the inclusion of any other scale in the vertex (beside $s$), this structure, although present, does not have the right IR limit.  However, this is not the end of the story, since there is a second kinematical configuration where the pole-like $1/s$ component becomes significant, and this involves the UV limit.  In fact, we are allowed to perform a large $s$ limit, in any direction away from the light cone, and observe the persistence of a $1/s$ component related to the anomaly. Notice that - differently from the case in which the two vector lines are on-shell - in this limit there is no redundancy between the longitudinal and transverse structure of the $L/T$ decomposition 
(the two structures are independent), and the $1/s$ behaviour is indeed a genuine (irreducible) part of the amplitude. 

Indeed, we can repeat the same analysis for the case in which at least one of the three scales $(m,s_1,s_2)$ is non-vanishing. Let's suppose, for instance, that only $m$ is non-zero. In this case we obtain (with $w_L(s_1,s_2,s,m^2)=W_L(0,0,s,m^2)$)

\bea
w_L(0,0,s,m^2)&=& -\frac{4 i}{s}\left[1+ \frac{m^2}{s}\log^2\left(\frac{a_3+1}{a_3-1}\right)\right],\\
w_T^{(+)}(0,0,s,m^2)&=& \frac{4 i}{s}\left[3+ \frac{m^2}{s}\log^2\left(\frac{a_3+1}{a_3-1}\right)-a_3 \log\left(\frac{a_3+1}{a_3-1}\right)\right],\\
w_T^{(-)}(0,0,s,m^2) &=& \tilde{w}_T^{(-)}(0,0,s,m^2) = 0, \qquad \qquad a_3=\sqrt{1-  \frac{ 4 m^2}{s}}.
\label{massive}
\eea
It is straightforward to verify that there is no residue for the $1/s$ pole term contained in $w_L$. This involves a cancellation between the two terms present in $w_L$, the constant and the logarithmic ($\sim \log^2$) term.

We conclude that the coupling of the pole in the infrared is controlled - in the absence of any other scale except $s$ in the diagram - by the $1/s$ component of $W_L$. This structure indeed saturates the anomaly. As soon as any other scale is generated, there is no IR coupling of this invariant amplitude, although it is formally present in the L/T decomposition. It is then clear that, if other scales are also present, we are still formally allowed to restore the Ward identities of the anomalous vertex by a subtraction of $W_L$ (which is what the GS mechanism does), but, by doing so, we have generated a vertex which is unique in its IR properties respect to any trilinear gauge vertex of the Standard Model. We refer to this situation as to an "oversubtraction" which can be potentially dangerous in the context of perturbative unitarity. This occurs whenever we move off-shell on the external lines (with $s_1$ or $s_2$ nonzero) or include a massive exchange in the loop, while still allowing an ordinary GS subtraction.  

A final comment, in this section, is due for the second (and independent) region where the $W_L$ contribution plays a role, which is the UV region. Notice that in the UV, being the external virtualities and mass negligible compared to the large value of $s$, we are again approaching the "pole dominance" typical of an IR ($m, s_1, s_2\sim 0$) amplitude. It is instructive to perform a large $s$ limit of the massive form factors given in (\ref{massive}), obtaining 
\bea
w_L &=& - \frac{ 4 \, i }{s} - \frac{ 4 \, i \, m^2}{s^2} \log \left( - \frac{s}{m^2}\right) + O (m^3), \\
w_T ^{(+)} (s,0,0,m^2 ) &=& \frac{12 \, i }{s} \, - \frac{4 \, i}{s}  \, \log \left( - \frac{s}{m^2}\right) + \frac{ 4 \, i \, m^2}{s^2} \left[ 2 +  \log \left(\frac{s^2}{m^4}\right) -  \log^2 \left( - \frac{s}{m^2}\right) \right] + O (m^3). \nn \\
\eea
The result above is susceptible of a simple intepretation. The anomalous contribution can be uniquely attributed to the pole in $W_L$, 
and the anomalous Ward identities are corrected by suppressed terms of the form $m^2/s$ which include logarithms of the same ratio. Differently from the $s\to 0$ case, in this limit of large $s$ there is no "overlap" between the two L/T tensor structures, and one can unambiguously attribute the anomalous contribution to $W_L$. This is the second - unequivocally distinct -  region where the 
anomalous $1/s$ contribution appears. It is somehow a misnomer, since there is no residue to compute in this case, but this contribution can still be called an "anomaly pole", since it is a manifestation of the anomaly and saturates the anomalous Ward identities as $s$ grows large. It is then clear which are the open issues typical of the mechanism of pole subtraction. If viewed as an asymptotic statement, we then should look for a completion of this mechanism. On the other hand, if we insist that the subtraction represents the only logical way to erase the anomalous variation of the action, then we are bound to face the issue of oversubtraction that we have mentioned before. 

\section{Quantifying the oversubtraction of an anomaly pole}
For the reasons mentioned above,  one can ask the question whether there is a completion of the GS mechanism - viewed as a pole subtraction - in order to avoid possible problems with the new (corrected) effective action in the infrared. 

The simplest possibility is to cancel identically the anomaly vertex and not just to restore its Ward identities under any kinematical configurations, which is what the pole subtraction does. We are going to do it using as a reference the ordinary cancellation via charge assignment, which allows to generate a complete unitary theory. 
 However, we will be separating the contribution to the cancellation which can be attributed to the exchange of the pseudoscalars, from the rest, with the residual interaction fixed by the condition of complete vanishing of the vertex. The residual terms, not included in the pole subtraction, could be attributed to the dynamics of the completion theory 
 (e.g. a string theory), but can be quantified in a definite form, as we are going to show, also in ordinary field theory.
 
Thus, let's consider a theory with a single chiral fermion with vector and axial-vector gauged interactions and the corresponding AVV diagram. A similar analysis can be done for the AAA diagram of the same model. 

We have seen that in this diagram any configuration - except for the on-shell case ($m, s_1,s_2=0$) of tuo $V$ lines - does not allow an identical cancellation of this diagram by a pole counterterm. It amounts, therefore, to an oversubtraction, as we have explained above. We denote this vertex by $W^{\lambda\mu\nu}(m, s,s_1,s_2)$ and 
using a standard Pauli-Villars regularization procedure, we subtract the same amplitude with a generic fermion of mass $M$ in the loop. 
We obtain, in a simplified notation, the regulated amplitude 
\beq
W_R=W(m) - W(M)
\eeq
which is obviously finite and satisfies ordinary Ward identities of the form 
\beq
k_{\lambda}W_R^{\lambda \mu\nu}= 2 m W^{\nu\lambda}(m) - 2 M W^{\nu\lambda}(M).
\eeq

Obviously, in a standard Pauli-Villars regularization one could send $M$ to infinity, recuperating the anomaly contribution from the 
$2 M W(M)$ term (up to a sign).  At this point we re-express each of the amplitudes in terms of a pole plus the transverse contributions obtaining 
\beq
W_R=\left(W_L(m, s_1,s_2)) + W_T(m, s_1,s_2)\right) - \left(W_L(M,s_1,s_2) + W_T(M,s,s_1,s_2)\right).
\label{reg1}
\eeq
Notice that each of $W_L(m, s_1,s_2)$ and $W_L(M,s_1,s_2)$ are made of an anomaly pole plus mass correction terms.  

Eq. (\ref{reg1}) can be decomposed in terms of $w_L$ and $w_T$, showing that $W_R$ is free of anomaly poles, leaving some extra contributions both in the $L$ and $T$ parts which are mass dependent. However, $W_R$ simplifies remarkably if the mass of the subtracted fermion is zero (M=0), since the anomaly diagram has no correction on the longitudinal structure $W_L$. In this specific case we obtain
\beq
W_R=\left(W'_L(m, s_1,s_2)) + W_T(m, s_1,s_2)\right) - W_T(s,s_1,s_2),
\label{reg2}
\eeq
where $W'_L$ denotes the $L$ component of the diagram for the physical fermion with the subtraction of the anomaly pole. 
The interpretation of equation (\ref{reg2}) is now obvious. Had we performed a pole subtraction on an $AVV$ diagram, $W(m,s_1,s_2)$, the result would have been given just by the first two terms in the round bracket, causing on oversubtraction. This is corrected by the second term $W_T(s,s_1,s_2)$ which performs the unitarization of the vertex at any scales. We stress once more that this unitarization is obtained from field theory arguments and does not necessarily correspond to the unitarization that a nonlocal completion theory, such as a string theory, should perform on the subtraction. 

We have gone through this argument to show that if the subtraction of a pole can be understood as a procedure which can be, eventually, unitarized in some way, then we can obviously give a coherent interpretation of the complete mechanism. This would allow us to attribute the subtraction of the pole term to one interaction, for instance to the exchange of an axion-ghost couple, while, at the same time, extra terms, not directly related to axionic contributions, would be involved in the extra correction. In the example that we have described, this extra term is given by $W_T(s,s_1,s_2)$, whose explicit expression, in this case, is known \cite{Armillis:2009sm}.

It is clear that there is a way out and a possible answer to the unitarization of the chiral anomaly pole, but it may not be so in the case of the trace anomaly.  It appears obvious that such a procedure is bound to fail in the trace anomaly case, unless extra contributions to the running of the beta function will manage to induce a conformal phase. In this respect, while a coherent formulation of a pole subtraction in superspace treats the trace and the chiral anomaly components of an anomaly supermultiplet equally, in practice one can't ignore the different nature of the two anomalies. This may pose severe constraints on the coupling of superanomaly multiplets to gravity, since the mechanism of cancellation of the anomaly, if realized by a pole subtractions in superspace, is not satisfactory. Pole-like contributions appear indeed both in the case of chiral and trace anomaly diagrams. However, the anomalous effective action generated by the insertion of arbitrary powers of the energy momentum tensor on correlators of gauge currents is far more involved.  It may not be completely saturated just by a pole to all orders, even in the weak field limit of the external gravitational field.

\section{Conclusions}
There is an incomplete understanding of the effective action which emerge at low energy from string theory and which involves a GS mechanism. It should be realized that this discussion is not just of formal nature, since it involves some issues which are of fundamental interest. First among them is the possible role played by the GS axion in the cosmology of the early Universe. The appearance of an axion is, in fact,  the crucial feature of 
the anomaly cancellation mechanism also in its realization in terms of a pole subtraction. The superspace formulation of the subtraction is not so obvious for K\"ahler anomalies, given the different nature of the chiral and conformal anomalies which are involved in combination in this subtraction.

Our analysis, clearly, is far from being conclusive, but it raises, we believe, some points which should motivate further discussions. Taken frontally, the subtraction of an anomaly pole to ensure the cancellation of some of the anomalies in a certain theory is
the correct thing to do. At the same time, however, it leaves some issues of consistency wide open. In fact, this approach could be possibly correct only in the on-shell case. By rewriting the nonlocal action into a local form, using a formulation with two extra degrees of freedom, one ghost and one axion, one indeed finds that the effective action breaks the Lorentz symmetry. In these effective actions the dynamical generation of the breaking is, in fact, rather economical. There is indeed a signal of vacuum instability in theories corrected by a pole subtraction, which seems to indicate that the ghost can be taken out of the physical spectrum, leaving for the rest a theory which could be potentially useful but in a nontrivial vacuum. 

This appears to be in agreement with the conclusions of 
\cite{ArkaniHamed:2003uy}, as indicated by Eq. (\ref{arka}).
Studies of gravity expanded around nontrivial background of ghosts are at the center of an increasing theoretical interest 
\cite{Holdom:2004yx, Thomas:2009uh} as are studies of the breaking of the Lorentz symmetry in brane models \cite{Mavromatos:2010ar,Mavromatos:2010pk}. Certainly, our comprehension of the vacuum structure of these theories on more physical grounds, especially in the presence of gravity multiplets, will probably require a big effort.

\centerline{\bf Acknowledgments}
 A.R. Fazio thanks the Physics Department of Universit\`{a} del Salento for the warm hospitality and acknowledges partial financial support from the Direcion de Investigacion (DIB) of Universidad Nacional de Colombia (sed Bogot{\`{a}}), project 9171  DIB Convocatoria Nacional de Investigacion 2009. The work of R. Armillis was supported by the European Union through the Marie Curie Initial Training Network ``UNILHC" PITN-GA-2009-237920.


\end{document}